\documentclass[10pt, twocolumn, prl]{revtex4-1}

\usepackage{natbib}

\pdfoutput=1
\usepackage{amsmath,amssymb,epsfig,color,bbold}
\usepackage{graphicx,caption}

\usepackage{bm}
\usepackage{latexsym}
\usepackage{amssymb}
\usepackage{epsfig,amsmath,graphicx}
\usepackage{listings,setspace}

\normalsize

\newcommand{\Lag}{\mathcal{L}}

\newcommand{\Br}{{\rm Br}}

\usepackage{tikz}
\usetikzlibrary{arrows,decorations.markings,decorations.pathmorphing}

\begin{document}

\begin{titlepage}

\begin{flushright}
FERMILAB-PUB-15-537-T\\

\today
\end{flushright}

\vspace{0.4cm}
\begin{center}
\Large\bf 
Creating the Fermion Mass Hierarchies with Multiple Higgs Bosons
\end{center}

\vspace{0.2cm}
\begin{center}
{\sc   Martin Bauer$^1$, Marcela Carena$^{2,3,4}$, Katrin Gemmler$^{2,5}$}\\
\vspace{0.4cm}
{$^1$ Institut f\"ur Theoretische Physik, Universit\"at Heidelberg, Philosophenweg 16, 69120 Heidelberg, Germany,\,\\
$^2$ Fermilab, P.O. Box 500, Batavia, IL 60510, USA,\, \\$^3$ Enrico Fermi Institute, University of Chicago, Chicago, IL 60637, USA, \\$^4$ Kavli Institute for Cosmological Physics,University of Chicago, Chicago, IL 60637, USA\,, \\$^5$ TUM Institute for Advanced Study, Technische Universit\"at M\"unchen, Arcisstra\ss e 21,
D-80333 M\"unchen, Germany.
}
\end{center}
\vspace{0.2cm}
\begin{abstract}
After the Higgs boson discovery, it is established that the Higgs mechanism explains electroweak symmetry breaking and generates the masses of all particles in the Standard Model, with the possible exception of neutrino masses. The hierarchies among fermion masses and mixing angles remain however unexplained. 
We propose a new class of two Higgs doublet models in which a flavor symmetry broken at the electroweak scale addresses this problem. The models are strongly constrained by electroweak precision tests and the fact that they produce modifications to Higgs couplings and flavor changing neutral currents; they are also constrained by collider searches for extra scalar bosons. The surviving models are very predictive, implying unavoidable new physics signals at the CERN Large Hadron Collider, \emph{e.g.} extra Higgs Bosons with masses $M < 700$ GeV. \end{abstract}
\maketitle

\end{titlepage}
\emph{Introduction.}
The observed flavor structure of quarks and leptons remains unexplained within the Standard Model (SM) of particle physics. Abelian flavor symmetries can create these hierarchies through different flavor charges of the SM fermions. This symmetry is broken by flavor charged SM singlet scalars,  so-called flavons $S_a$, when they acquire vacuum expectation values $\langle S_a \rangle=f_a$. Yukawa couplings are then generated through higher order operators that involve powers of the expansion parameters $\varepsilon_a = f_a/\Lambda_a$, where $\Lambda_a$ are the scales at which the New Physics (NP) sets in.
The Froggatt-Nielsen model  \cite{Froggatt:1978nt}
implements this idea with a
single abelian flavor group $U(1)_F$. The effective Yukawa couplings for the down-type fermions are then given by
\begin{equation}
y^d_{ij}\left(\frac{S}{\Lambda}\right)^{n^d_{ij}}\!\!\bar Q_i \Phi\, d_{R_j} \quad \rightarrow \quad Y^d_{ij}\,\bar Q_i \Phi\, d_{R_j}\,,
\end{equation}
with
$Y_{ij}^d\propto \varepsilon^{n^d_{ij}}\, y_d^{ij}, \, (i,j =1,2,3)$, in which the $y_d^{ij}$ are structureless order one coefficients. The exponents  $n^d_{ij}$ depend on the flavor charges of the $SU(2)_L$ quark doublets $Q_i=( u_{L_i}, d_{L_i})^T$ and singlets $d_{R_j}$, and the Higgs $SU(2)_L$ doublet $\Phi$. Similar expressions hold for the up-type fermions. 
Given that the Yukawa couplings are dimensionless, this mechanism does neither fix $\Lambda$ nor the scale of flavor symmetry breaking $f$. As a result the relevant scales can be arbitrarily high, rendering this idea unverifiable.  If instead the SM Higgs plays a role in the flavor symmetry breaking, the flavor symmetry breaking scale is replaced by the electroweak scale $v = 246 $ GeV. This ansatz has been proposed in the literature with the SM singlet combination $\Phi^\dagger \Phi$ acting as the flavon \cite{Babu:1999me, Giudice:2008uua}. In this case the expansion parameter becomes $\varepsilon \approx v^2/\Lambda^2$, thereby predicting the NP scale $\Lambda \propto v/\sqrt{\varepsilon}$. However, the shortcoming of this idea is that $\Phi^\dagger \Phi$ cannot carry a flavor quantum number. Furthermore, measurements of Higgs couplings at the LHC exclude this ansatz.   
In this letter we introduce a new class of models with two Higgs doublets 
$H_1$ and $H_2$, such that both may carry a non-trivial flavor charge and act jointly as the flavon $S$. The Higgs boson properties single out a preferred region of parameter 
space, that is unique to this class of two Higgs doublet flavor models (2HDFM).
The flavor off-diagonal couplings as well as the properties of the new scalars lead to striking signals that can be discovered at the Large Hadron Collider (LHC). 

\vspace{2mm}
\emph{Electroweak Scale Flavor Models.}
The general set-up is defined by effective Yukawa couplings 
\begin{align}\label{eq:yuks}
\Lag_Y&\ni  y^u_{ij}\bigg[\!\frac{H_1^{\phantom{\dagger}} H_2^{\phantom{\dagger}}  }{\Lambda^2}\!\bigg]^{\!\eta^u_{ij}}\!\!\bar Q_i H_1\, u_{R_j} +\tilde y^u_{ij}\bigg[\!\frac{H_1^{\phantom{\dagger}} H_2^{\phantom{\dagger}}  }{\Lambda^2}\!\bigg]^{\!\xi^u_{ij}}\!\!\bar Q_i \tilde H_2\, u_{R_j} \notag \\
&\hspace{-.35cm}+y^d_{ij}\bigg[\!\frac{H_1 H_2 }{\Lambda^2}\!\bigg]^{\!\eta^d_{ij}}\!\!\bar Q_i H_2\, d_{R_j}+\tilde y^d_{ij}\bigg[\!\frac{H_1 H_2 }{\Lambda^2}\!\bigg]^{\!\xi^d_{ij}}\!\!\bar Q_i \tilde H_1\, d_{R_j},
\end{align}
in which $\tilde H_{1, 2}\equiv  i\sigma_2 H_{1, 2}$.
The SM gauge singlet 
$H_1 H_2$ 
carries a flavor charge, and the number of insertions $\eta^u_{ij}, \xi^u_{ij}$ and $\eta^d_{ij},\xi^d_{ij}$ are fixed in order to reproduce the observed fermion mass hierarchies. Depending on the flavor charge assignment of the quarks and Higgs fields, some of the above operators should be written replacing the SM gauge singlet $H_1 H_2$ by  $H_1^\dagger H_2^\dagger$. The coefficients $y^u_{ij}, \tilde y^u_{ij}$ and $y^d_{ij}, \tilde y^d_{ij}$ are assumed to be structureless and all of order one.  We consider two cases: If $\eta^{u,d}_{ij} < \xi^{u,d}_{ij}$, the Yukawa sector \eqref{eq:yuks} at leading order in $\varepsilon$ reads \cite{Bauer:2015fxa}
\begin{align}\label{eq:yuksII}
\hspace{-1mm}\Lag_Y^\mathrm{II}&\!\ni\!y^u_{ij} \bigg[\!\frac{H_1^{\phantom{\dagger}} H_2^{\phantom{\dagger}}  }{\Lambda^2}\!\bigg]^{\!\eta^u_{ij}}\!\!\bar Q_i H_1\, u_{R_j}\!+\!y^d_{ij}\bigg[\!\frac{H_1 H_2 }{\Lambda^2}\!\bigg]^{\!\eta^d_{ij}}\!\!\bar Q_i H_2\, d_{R_j},\!
\end{align}
which corresponds to a two Higgs doublet model (2HDM) of type II in the limit $\eta^u_{ij}\rightarrow 0$ and $\eta^d_{ij}\rightarrow 0$ \cite{Branco:2011iw}. For $\eta^u_{ij} < \xi^u_{ij}$, but $\eta^d_{ij} > \xi^d_{ij}$, the Yukawa couplings at leading order in $\varepsilon$ instead read
\begin{align}\label{eq:yuksI}
\hspace{-1mm}\Lag_Y^\mathrm{I}\!\ni\!y^u_{ij} \bigg[\!\frac{H_1^{\phantom{\dagger}} H_2^{\phantom{\dagger}}  }{\Lambda^2}\!\bigg]^{\!\eta^u_{ij}}\! \!\bar Q_i H_1\, u_{R_j}\!+\!\tilde y^d_{ij}\bigg[\!\frac{H_1^\dagger H_2^\dagger }{\Lambda^2}\!\bigg]^{\!\xi^d_{ij}}\!\!\bar Q_i \tilde H_1\, d_{R_j},\!
\end{align}
which corresponds to a 2HDM of type I in the limit of vanishing flavor charges $\eta^u_{ij}\rightarrow 0$ and $\xi^d_{ij}\rightarrow 0$. In the following we consider cases \eqref{eq:yuksII} and \eqref{eq:yuksI} and refer to them as 2HDFM type II and type I, respectively. Additional options for which $\eta^q_{ij} < \xi^q_{ij}$ only for some entries $ij$ will be discussed elsewhere. We concentrate on the quark sector and, with the exception of tau leptons that impact Higgs decays, we reserve the modeling of the lepton sector for a later study.
The flavor breaking scale is set by the vacuum expectation values of the two Higgs doublets $  \langle H_{1, 2} \rangle\equiv v_{1, 2}$,  
\begin{equation}\label{eq:eps}
\varepsilon = \frac{v_1 v_2}{\Lambda^2}=\frac{\tan \beta}{1+\tan^2 \beta}\frac{v^2}{2\Lambda^2},
\end{equation}
where $\tan \beta \equiv v_1/v_2$, and larger values of $\tan\beta$ yield lower NP scales $\Lambda$.  
Furthermore, the effective Yukawa couplings in the weak eigenbasis are defined as 
$Y^{u}_{ij}= y^u_{ij}\,\varepsilon^{\eta_{ij}^u}$\,, $ Y^{d, \mathrm{I}}_{ij}= \tilde y^d_{ij}\,\varepsilon^{\xi_{ij}^d}$,  and $Y^{d, \mathrm{II}}_{ij}= y^d_{ij}\,\varepsilon^{\eta_{ij}^d}$. After rotation to the quark mass eigenbasis, we can write $m_{q_i}\propto m_t\,\varepsilon^{n_{q_i}}$, with $q=u,d$ and $\varepsilon \equiv m_b/m_t$, such that the exponents 
$n_{u_i}= n_u, n_c, n_t$ and $n_{d_i}=n_d, n_s, n_b$ are fixed to reproduce the quark masses  
\begin{equation}\label{eq:nqis}
n_t=0,\, n_b=n_c=1,\, n_s=2,\, n_d= n_u=3 \,.
\end{equation}
The structure of the Cabbibo-Kobayashi-Maskawa (CKM) fermion mixing matrix imposes additional constraints on the flavor charges. We consider two different choices $V_a$ and $V_b$, with CKM entries $V^{12}_a=1$, $V^{13}_a=V^{23}_a=\varepsilon$ and $V^{12}_b=V^{13}_b=V^{23}_b=1$.
The structure $a$ has the largest CKM hierarchy build in, while $\mathcal{O}(1)$ hierarchies rely on different values of the fundamental Yukawas $y^u_{ij}, y^d_{ij}$ or $\tilde y^d_{ij}$. For structure $b$ all CKM hierarchies rely on cancelations between $\mathcal{O}(1)$ fundamental Yukawa couplings.
In the type II 2HDFM both structures are possible \eqref{eq:yuksII}, while in the type I 2HDFM \eqref{eq:yuksI}, only the structure $b$ can be realized, otherwise some of the $\eta^d_{ij} < \xi^d_{ij}$. \\
\begin{figure*}[t]
\begin{center}
\begin{tabular}{ccc}
\includegraphics[width=1\textwidth]{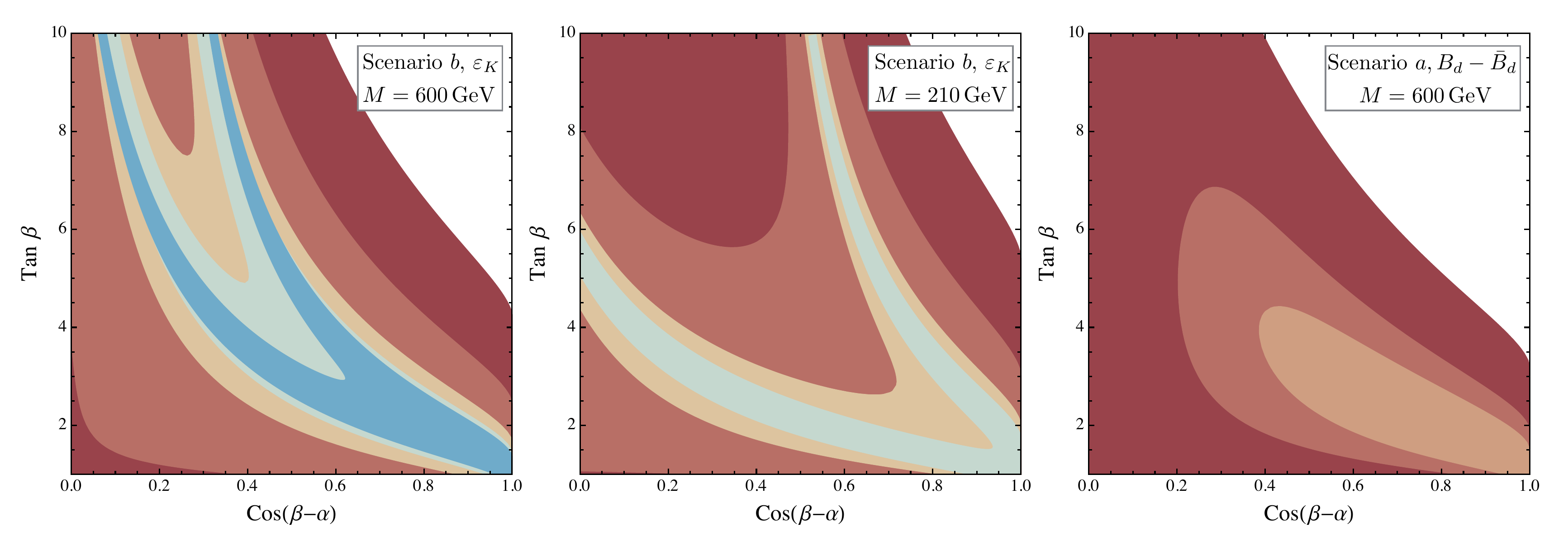}
\end{tabular}
\end{center}
\caption{
\label{fig:flavplot}Flavor constraints from $\epsilon_K$ for the flavor structure $b$ and $M=600$ GeV (left) and for the flavor structure $b$ and $M=210$ GeV (center), as well as from $C_{B_d}$ for the flavor structure $a$ and $M=600$ GeV (right). The color coding corresponds to suppressions of the flavor structures by factors $0.1$ (dark red), $0.2$ (light red), $0.4$ (orange), $0.6$ (light green) and $0.8$ (light blue). See text for details. 
}
\end{figure*}

\vspace{1mm}
\emph{Higgs Couplings.}
The Higgs sector contains two neutral scalar mass eigenstates $h, H$ with a mixing angle $\alpha$, and one pseudoscalar $A$ and one charged scalar $H^\pm$, both with mixing angle $\beta$. The lighter scalar mass eigenstate $h$ is identified with the $125$ GeV resonance observed at the LHC. 
Scalar self-couplings as well as couplings to gauge bosons $V=W^\pm, Z$ are the same as in generic 2HDMs.
In particular,
the couplings between two gauge bosons and a single neutral scalars $\varphi=h,H$ are given by
 $g_{\varphi VV}=\kappa^\varphi_V \, 2\,m_V^2/v$, with 
$\kappa^h_V=\sin(\beta-\alpha)\,, \, \kappa^H_V=\cos (\beta-\alpha)$. The structures of the effective Yukawa matrices are determined by powers of $\varepsilon$. After rotation to the Higgs and quark mass eigenbases they induce couplings of the scalar Higgs bosons to quarks $u_i = u_{L_i}+ u_{R_i}$ and $ d_i= d_{L_i}+ d_{R_i}$,
\begin{equation}\label{eq:newlag}
\Lag= g_{\varphi u_{L_i}  u_{R_j} }\, \varphi \,\bar u_{L_i} u_{R_j}+ g_{\varphi d_{L_i}d_{R_j} }\, \varphi \,\bar d_{L_i} d_{R_j}+h.c.\,.
\end{equation}
The flavor-diagonal couplings read
\begin{align}\label{eq:diagcoup}
\! \! g_{\varphi q_{L_i}q_{R_i}}\!= \kappa^\varphi_{q_i}\, \frac{m_{q_i}}{v}=\Big(g^{\varphi}_{q_i}(\alpha,\beta)+n_{q_i}\, f^\varphi(\alpha, \beta)\!\Big)\frac{m_{q_i}}{v},
\end{align}
with the flavor universal functions $f^\varphi$ given by
\begin{align}\label{eq:fFs}
f^h(\alpha,\beta)=\frac{c_\alpha}{s_\beta}-\frac{s_\alpha}{c_\beta}\,,\qquad 
f^H(\alpha,\beta)=\frac{c_\alpha}{c_\beta}+\frac{s_\alpha}{s_\beta}\,,
\end{align}
and the flavor dependent functions defined as
\begin{equation}
\begin{tabular}{ccccc}
&$g^h_{u_i}$&$g^h_{d_i}$&$g^H_{u_i}$&$g^H_{d_i}$\\[2pt] \hline 
Type I& $\frac{c_\alpha}{s_\beta} $&$ \frac{c_\alpha}{s_\beta}$&$\frac{s_\alpha}{s_\beta}$&$\frac{s_\alpha}{s_\beta}$ \\
Type II&$ \frac{c_\alpha}{s_\beta}$ &$- \frac{s_\alpha}{c_\beta}$&$\frac{s_\alpha}{s_\beta}$&$\frac{c_\alpha}{c_\beta}$ \\
\end{tabular}\,,
\label{eq:coupfac}
\end{equation}
where $s_{\alpha} = \sin \alpha$ and $c_{\alpha} = \cos \alpha$ and similarly for $\alpha \leftrightarrow \beta$.
 The flavor off-diagonal couplings in \eqref{eq:newlag} read ($i \neq j$)
 \begin{align}\label{eq:foff}
g_{\varphi q_{L_i}q_{R_j}}=  f^\varphi(\alpha, \beta)\,G_{ij}\,\equiv f^\varphi(\alpha, \beta)\left(\mathcal{Q}_{ij}\frac{m_{q_j}}{v}-\frac{m_{q_i}}{v}\mathcal{B}_{ij} \right)\,,
\end{align}
where  $\mathcal{Q}$, and $\mathcal{B}=\mathcal{D}, \mathcal{U}$ are symmetric matrices with off-diagonal entries of $\mathcal{O}(\varepsilon)$ or smaller. The $\varepsilon$-dependence is uniquely fixed by the conditions from the quark masses \eqref{eq:nqis} and from the CKM matrix, while coefficients depending on the fundamental Yukawas $y^u_{ij}, y^d_{ij}$ or $\tilde y^d_{ij}$ can lead to further suppressions. The $\varepsilon$-dependence is given by
  \begin{align}
\label{eq:fstruc}
\begin{tabular}{c|ccc|ccc}
&$\mathcal{Q}^a_{ij}$&$\mathcal{U}^a_{ij}$&$\mathcal{D}^a_{ij}$&
$\mathcal{Q}^b_{ij}$&$\mathcal{U}^b_{ij}$&$\mathcal{D}^b_{ij}$
\\
\hline
$(1,2)$&$\varepsilon^2$&$\varepsilon^2$&$\varepsilon^{\phantom{2}}$&$0$&$\varepsilon^2$&$\varepsilon^{\phantom{2}}$   \\
$(1,3)$&$\varepsilon^{\phantom{2}}$&$\varepsilon^4$&$\varepsilon^{\phantom{2}}$
&$0$&$\varepsilon^3$&$\varepsilon^2$\\
$(2,3)$&$\varepsilon^{\phantom{2}}$&$\varepsilon^2$&$\varepsilon^2$&$0$&$\varepsilon^{\phantom{2}}$&$\varepsilon^{\phantom{2}}$
\end{tabular}\,.
\end{align}
The couplings of the pseudoscalar to quarks are given by $g_{Aq_{L_i}q_{R_j}}= i \, g_{Hq_{L_i}q_{R_j}}\vert_{c_\alpha\rightarrow s_\beta, s_\alpha\rightarrow c_\beta} $, with the corresponding flavor universal function $f^A(\beta)=\tan \beta+ \cot \beta$. The trilinear charged Higgs--quark pair vertex is the same as in the 2HDM of type I or II, respectively. A distinguishing feature of the 2HDFMs is that the same function $f^\varphi(\alpha,\beta)$ controls both the flavor-changing couplings and the departure of the flavor diagonal Higgs-fermion couplings from a 2HDM type I or II.
In the decoupling limit, where all additional Higgs bosons are sufficiently heavy, there is alignment: $\alpha=\beta-\pi/2$. This results in $f^h(\beta-\pi/2,\beta)=2$ and hence $g_{hb_Lb_R}=3 m_b/v$, which is experimentally excluded \cite{Babu:1999me,Giudice:2008uua}. Therefore, to avoid this limit, the new scalars cannot be arbitrarily heavy, and should be accessible at the LHC.\\
\begin{figure*}[t!]
\begin{center}
\begin{tabular}{ccc}
\hspace{-.3cm}\includegraphics[width=1.03\textwidth]{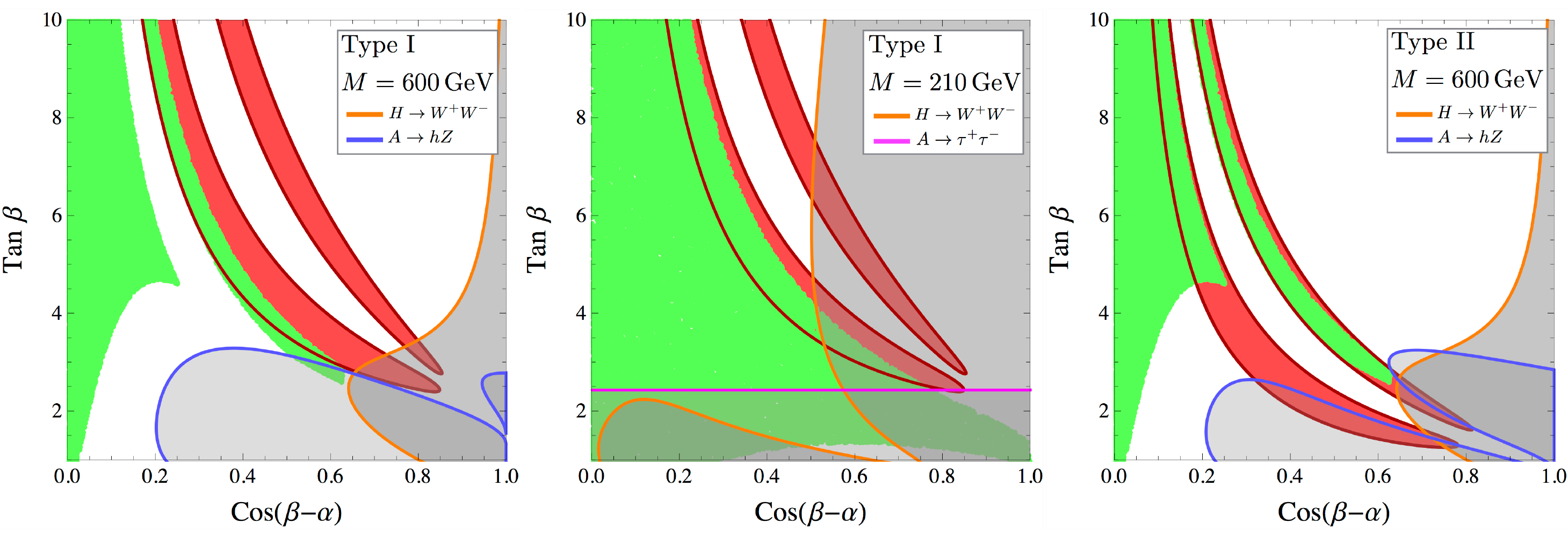}
\end{tabular}
\end{center}
\caption{\label{fig:finplot} 
Constraints in the $\cos (\beta-\alpha) - \tan \beta$ plane for the type I 2HDFM and $M=600$ GeV ($M=210$ GeV) in the left (center) panel as well as for type II 2HDFM and $M=600$ GeV in the right panel. The areas preferred by the $2\sigma$ Higgs global fits are shaded red and those
 allowed by EWPM bounds are shaded green. Areas excluded by collider bounds are shaded grey.
See text for details.}
\end{figure*}

\emph{Flavor Constraints.}
 Due to the flavor off-diagonal couplings of the scalars in \eqref{eq:foff},  FCNCs are mediated by $s$-channel tree-level exchange of $h, H$ and $A$. The strongest constraints arise from $\Delta F=2$ processes. The NP tree level contributions to these processes are captured by the effective Hamiltonian
\begin{equation}\label{eq:heffdf2}
\mathcal{H}_\mathrm{NP}^{\mathrm{tree}}=
C_2^{ij} \,\mathcal{O}^{ij}_2+\tilde C_2^{ij}\,\tilde{ \mathcal{O}}^{ij}_2+ C_4^{ij}\,\mathcal{O}^{ij}_4 
\end{equation}
in which $i,j$ are flavor indices and $\mathcal{O}_2^{ij}= (\bar q^i_R \, q^j_L)(\bar q^i_R \, q^j_L)$, 
$\tilde{\mathcal{O}}_2^{ij}= ( \bar q^i_L \, q^j_R)( \bar q^i_L \, q^j_R)$, and 
 $\mathcal{O}_4^{ij} = ( \bar q^i_R \, q^j_L)\, ( \bar q^i_L \, q^j_R)$.
The Wilson coefficients read
\begin{align}\label{eq:C2}
\hspace{-.18cm}C^{ij}_x \!= \!
-\frac{c_x^{ij}}{v^2}\,
\bigg\{\!\frac{f^h(\alpha,\beta)^2}{m_h^2}+\frac{f^H(\alpha,\beta)^2}{M_H^2}\pm \frac{f^A(\beta)^2}{M_A^2}\!\bigg\}\,,
\end{align}
with $C_x^{ij}= C_4^{ij}$ for the positive relative sign and $C_x^{ij}= C_2^{ij}, \tilde C_2^{ij} $ for the negative sign. In \eqref{eq:C2}, the flavor dependent part is defined as
\begin{align}
c_2^{ij}\equiv (G^\ast_{i j})^2 \, , \, \, \tilde c_2^{ij} \equiv (G_{ji} )^2\,,\,\,
c_4^{ij} \equiv G^\ast_{i j} G_{ji } / 2 \,.
\end{align}
Therefore, the size of the Wilson coefficients \eqref{eq:C2} is controlled by the Higgs boson masses, the flavor universal functions $f^\varphi(\alpha,\beta)$, and the explicit flavor-dependent coefficients $G_{ij}$ suppressed through their dependence on the fermion masses and the structures in \eqref{eq:fstruc}. 
These structures define a hierarchy between the NP contributions to $K -\bar K$, $B_{d/s}-\bar B_{d/s}$ and $D-\bar D$ mixing. Additional contributions arise at loop-level, and are dominated by charged Higgs and $W^\pm$ exchange. These radiative corrections most importantly affect the Wilson coefficient $C_1\mathcal{O}_1=C_1(\bar q^i_L\gamma_\mu q^j_L)(\bar q^i_L\gamma^\mu q^j_L)$ already present in the SM, however, they are additionally suppressed by a parametric dependence $C_1^\mathrm{NP}\propto \cot^2 \beta$ for both type I and type II 2HDFMs. Other loop contributions are subleading.
The contributions to $K-\bar K$ mixing from both flavor scenarios are of similar order, and $\epsilon_K$ yields the strongest constraint for the structure $b$ . The left and center panel of Fig.  \ref{fig:flavplot} show contours of 
\begin{equation}
C_{\epsilon_K}= \frac{\mathrm{Im} \, \langle K^0 | \mathcal{H}_\mathrm{SM} + \mathcal{H}_\mathrm{NP} |\bar K^0\rangle}{ \mathrm{Im} \langle K^0 | \mathcal{H}_\mathrm{SM} |\bar K^0\rangle} =1.05^{+0.36}_{-0.28} \,, 
\end{equation}
in agreement with the experimental constraint \cite{Bona:2007vi},
for masses $M=M_H=M_A=M_{H^+}= 210$ GeV (left) and $M=600$ GeV (center), respectively. The constraints from $B_{d}-\bar B_{d}$ mixing are the most stringent for the structure $a$, and we show contours of
 \begin{equation}
C_{B_d}=\left| \frac{ \langle B^0_d | \mathcal{H}_\mathrm{SM} + \mathcal{H}_\mathrm{NP} |\bar B^0_d\rangle}{  \langle B_d^0 | \mathcal{H}_\mathrm{SM} |\bar B^0_d\rangle}\right|=1.07^{+0.36}_{-0.31},
\end{equation}
in agreement with the experimental bound \cite{Bona:2007vi} in the right panel of Fig.  \ref{fig:flavplot}. The color coding shows a suppression of the structures \eqref{eq:fstruc} by factors  
$0.1$ (dark red), $0.2$ (light red), $0.4$ (orange), $0.6$ (light green) and $0.8$ (light blue), which are the result of accidental cancellations in the coefficients depending on $y^q_{ij}$ and $\tilde y^q_{ij}$. The light blue, light green and orange regions are therefore preferred by flavor constraints, while the light and dark red regions imply a moderate tuning of a priori random $\mathcal{O} (1)$ parameters. In scenario $b$ only the structure $\mathcal{Q}$ or $\mathcal{D}$ are revelant, while in scenario $a$, coefficients in front of both structures $\mathcal{Q}$ and $\mathcal{D}$ in \eqref{eq:fstruc} need to be simultaneously small and hence the tuning is more severe. A numerical study for scenario $a$ reveals, that therefore only $16\%, 4\%, 1\%$ of parameter sets are allowed in the orange, light red, red regions, respectively. \\
Constraints from $B_s-\bar B_s$ and $D - \bar D$ mixing turn out to be less stringent for both flavor structures.  One important difference between type I and II 2HDFMs appears in NP effects in $b\rightarrow s \gamma$. Type I models have an additional $\cot^2\beta$ suppression in the contribution from the dominant charged Higgs penguin diagram, which for $\tan \beta \gtrsim 2$ relaxes the otherwise important bound on the charged Higgs mass $M_{H^\pm} > 360 $ GeV at $3\sigma$, effective in type II models \cite{Khachatryan:2014wca, Misiak:2015xwa}.

\vspace{2mm}
\emph{Electroweak precision tests, perturbativy and unitarity.} 
In 2HDMs, arbitrarily large masses of the extra scalars imply $\cos (\beta-\alpha)=0$ or non-perturbative quartic couplings. We implement the perturbativity condition for each quartic coupling individually and obtain that values of $\cos (\beta-\alpha)\gtrsim 0.1$ and $\tan \beta \lesssim 10$ require extra scalar masses $\lesssim 700$ GeV.
The stability of the scalar potential, the unitarity of the S matrix and contributions to the oblique $S$, $T$ and $U$ parameters further restrict the allowed region in the $\cos(\beta-\alpha) - \tan \beta$ plane and limit the permitted mass splittings between the heavy scalar, pseudoscalar and charged Higgs boson masses. 
In Fig. \ref{fig:finplot}, we show in green the regions allowed by all the above constraints for the type I 2HDFM and $M=600$ GeV ($M=210$ GeV) in the left (center) panel, as well as for type II 2HDFM and $M=600$ GeV in the right panel. Fig. \ref{fig:finplot} includes perturbativity and unitarity conditions, as well as bounds from electroweak precision measurements (EWPM) at the $2\sigma$ level, based on the Two-Higgs-Doublet Model Calculator \cite{Eriksson:2009ws} and the values provided by the Gfitter analysis \cite{Baak:2014ora}.  
 \\

\vspace{2mm} 
\paragraph{Collider Phenomenology.}
We consider a global fit to the Atlas and CMS Higgs signal strength measurements as well as LHC constraints from searches for heavy resonances based on the scalar couplings presented in the Higgs coupling section.  For the Higgs-tau lepton coupling, we assume the analogous of \eqref{eq:diagcoup} with $n_\tau=1$. \\
Based on ATLAS data, Fig.  \ref{fig:finplot} shows the results of the $2\sigma$ global fit \cite{ATLAS:combinedHiggsdata} as red shaded regions for the 2HDFM type I (left and center panel) and type II (right panel).  As expected, the regions preferred by the global fit in both cases are far from the decoupling limit. Interestingly, two branches corresponding to positive (lower branch) and negative (upper branch) signs of the down-type Yukawa couplings are allowed.  A similar fit using CMS data \cite{CMS:Moriond} is less constraining. 
\\
Searches for heavy Higgs bosons with masses below $600$ GeV at the LHC are particularly promising. We present two scenarios with common heavy scalar masses $M=210$ GeV and $M=600$ GeV. For the latter we discuss only the type I 2HDFM, since type II ist strongly constrained by $b \rightarrow s \gamma$, requiring $M_{H^+}\gtrsim 360$ GeV.
We generate the gluon-fusion production cross section at next-to-leading order using HIGLU \cite{Spira:1995mt}, including the contributions of the bottom quark loop. For the vector-boson fusion production cross section we use the values quoted in \cite{Dittmaier:2011ti, Heinemeyer:2013tqa}. Further, we take the leading order expressions for the partial decay widths \cite{Djouadi:1995gv}. For both production and decay processes, we adjust the couplings to our model. 
The neutral CP-even Higgs $H$ decays dominantly to vector gauge bosons and depending on the $\cos(\beta-\alpha)-\tan \beta$ region can be most efficiently produced via the gluon fusion or the vector boson fusion processes. 
We compute $\sigma (pp \to H +X) \times \Br(H \to VV)/(\sigma (pp \to H+X) \times \Br(H \to VV))_{\text{SM}}$ and, based on the CMS analysis \cite{Khachatryan:2015cwa}, show the excluded parameter space shaded grey with orange contours in Fig.  \ref{fig:finplot}.  
The subdominant $H \to hh$ decay mode provides no additional bounds at present. 
Heavy CP-odd Higgs bosons are sought for by the CMS \cite{Khachatryan:2015lba} and {ATLAS} experiments \cite{Aad:2015wra} through $\sigma(gg \to A)\times \Br(A \to hZ)$ with a subsequent decay of $h\to b\bar b$ or $h \to \tau^- \tau^+$. 
For heavy (pseudo)scalars, $M_{H, (A)}\gtrsim 500$ GeV and away from alignment, finite width effects become important ,and following the CMS analysis \cite{CMS:2013yea}, we include finite width effects in $A \to hZ$. For $M=600$ GeV we show the corresponding excluded regions shaded grey with blue contours in the left panel (type I) and right panel (type II) of Fig. \ref{fig:finplot}. 
For $M=210$ GeV the $A\rightarrow hZ$ channel is kinematically forbidden and the decay mode $A\rightarrow \tau^+\tau^-$ provides the strongest bound from pseudoscalar searches. The corresponding excluded parameter space is shaded grey with a magenta contour in the center panel of Fig. \ref{fig:finplot}. The presently allowed region in the $\cos(\beta-\alpha)-\tan\beta$ plane in Fig. \ref{fig:finplot} is shown by the overlap of the red and green areas that is not covered by grey shading. \\

\emph{Conclusion.} 
We present a new class of models with two Higgs doublets to create the fermion mass hierarchies at the electroweak scale. The textures of the Yukawa couplings are a result of an abelian flavor symmetry that only allows renormalizable Yukawa couplings of the top quark to the Higgs bosons. All other Yukawa couplings are generated by higher dimensional operators that produce hierarchical entries of the Yukawa matrices, explaining the observed quark masses and mixing angles.
We study two different realizations of this class of models that differ in the roles the two Higgs doublets play in generating up- and down-quark masses and the CKM matrix. Flavor observables, LHC Higgs signal strength measurements, EWPMs, unitarity and perturbativity bounds, as well as collider searches for new scalar resonances 
result in precise predictions
for the parameters of these 2HDFMs. The beauty of solving the flavor problem at the electroweak scale is that striking signals such as correlated departures from SM Higgs couplings, as well as additional Higgs bosons with masses $< 700$ GeV must be observed at the LHC.\\

\begin{acknowledgments}
\paragraph{Acknowledgements.}\hspace{-.2cm} We thank Adrian Carmona, Lawrence Hall and Alexey Petrov for interesting discussions. 
MB acknowledges the support of the Alexander von Humboldt Foundation. Fermilab is operated by Fermi Research Alliance, LLC under Contract No. DE-AC02- 07CH11359 with the United States Department of Energy. KG acknowledges support by the Deutsche Forschungsgemeinschaft (DFG), grant number GE 2541/2-1. 
\end{acknowledgments}

\bibliography{hff.bib}

\end{document}